\documentclass[utf8]{webofc}
\usepackage[varg]{txfonts}   
\usepackage[per-mode=symbol,range-phrase=\text{--},range-units=single,separate-uncertainty=true,detect-shape,binary-units=true]{siunitx}
\newcommand{\CC}{C\nolinebreak\hspace{-.05em}\raisebox{.4ex}{\tiny\bf ++}}

\newcommand{\rdfnode}[1]{\emph{#1}}
\newcommand{\Define}{\rdfnode{Define}}
\newcommand{\HistoOneD}{\rdfnode{Histo1D}}
\newcommand{\Filter}{\rdfnode{Filter}}
\newcommand{\Selection}{\emph{Selection}}
\newcommand{\Plot}{\emph{Plot}}
\newcommand{\RDF}{RDF}
\newcommand{\eg}{e.g.\ }
\newcommand{\etc}{etc.\ }

\usepackage{hyperref}

\usepackage{fancyvrb}
\usepackage{color}
\input{pygm_defs.tex}
\begin{document}
\hfill{}CP3-21-05%
\title{Readable and efficient HEP data analysis with bamboo}
%
%

\author{\firstname{Pieter} \lastname{David}\inst{1}\fnsep\thanks{\email{pieter.david@cern.ch}}}

\institute{Universit\'{e} catholique de Louvain, Louvain-la-Neuve, Belgium}

\abstract{%
  With the LHC continuing to collect more data and experimental analyses becoming
  increasingly complex, tools to efficiently develop and execute these analyses
  are essential.
  The bamboo framework defines a domain-specific language, embedded in python,
  that allows to concisely express the analysis logic in a functional style.
  The implementation based on ROOT's RDataFrame and cling \CC{} JIT compiler
  approaches the performance of dedicated native code.
  Bamboo is currently being used for several CMS Run 2 analyses that rely on
  the NanoAOD data format, which will become more common in Run 3 and beyond,
  and for which many reusable components are included, but it provides many
  possibilities for customisation, which allow for straightforward adaptation
  to other formats and workflows.
}
\maketitle
\section{Introduction}
\label{intro}

As the LHC moves to a third run, and then a high-luminosity phase,
where the gain in physics reach mostly comes from the increase in
the size of the data samples than from an increase in centre-of-mass energy,
the focus of experimental measurements shifts towards improving the precision.
Data collected over longer periods of time are jointly analyzed,
and calibrations, corrections, and data analysis techniques become
increasingly fine-grained and sophisticated.
This also has an impact on data analysis workflows, where there is a trend
towards more compact and standardized data formats, \eg the introduction of
the NanoAOD format in the CMS collaboration~\cite{Rizzi:2019rsi},
which stores about \SI{1}{\kilo\byte} of information per collision or simulated event,
and is foreseen to cover the needs of a major fraction of the measurements
and searches.

With the increasing complexity of the experimental analyses,
the computer code to perform them easily becomes inefficient,
if simplicity and readability is given priority,
or inflexible and error-prone due to the need to efficiently
process large amounts of data.
The bamboo\footnote{see also \url{https://gitlab.cern.ch/cp3-cms/bamboo} and
\url{https://cp3.irmp.ucl.ac.be/~pdavid/bamboo/}
for the source code repository and documentation, respectively}
package avoids this trade off by providing a flexible python framework
with a high-level interface to construct computation graphs that can efficiently
be executed by RDataFrame~\cite{enrico_guiraud_2017_260230} (\RDF{}),
a declarative columnar interface for data analysis in the ROOT
framework~\cite{Brun:1997pa,fons_rademakers_2020_3895852}.
The core of the solution is an embedded domain-specific language,
an analysis description language~\cite{Sekmen:2020vph} with elements of
functional-style programming.
The ability of ROOT and \RDF{} to include, compile, and call arbitrary \CC{}
code through cling~\cite{cling}, with automatic python bindings through
PyROOT~\cite{lavrijsen2016high}, is used to provide a set of features
that is currently complete enough to perform several CMS analyses
in the context of top quark production measurements,
searches for rare production modes of the Brout-Englert-Higgs scalar boson,
and searches for additional electroweak-scale scalar bosons,
on the full data sample collected by the CMS experiment
during the second run of the LHC (2015--2018).

The following sections provide a description of the embedded domain-specific
language to construct derived per-event quantities and its implementation
(Section~\ref{sec:eventviewexpressions}), the construction of execution graphs
through \RDF{} (Section~\ref{sec:selectionplotsyst}),
the interface for dedicated tools to implement on-demand corrections
(Section~\ref{sec:cpptools}), and the structure of the framework to maximise
the reuse of code that is not part of the analysis logic
(Section~\ref{sec:framework}).
Finally ideas for future developments (Section~\ref{sec:futuredirections})
and a conclusion (Section~\ref{sec:conclusion}) are presented.

\section{Event view and constructing derived quantities}
\label{sec:eventviewexpressions}

The defining characteristic of a declarative columnar interface like \RDF{}
is that instead of writing code that is executed for every event, or entry in
the dataset, a set of transformations and operations is defined based on the
columns, before reading any of the data, which are then efficiently executed
by the backend.
Since \RDF{} allows to pass \CC{} strings that are then just-in-time compiled,
filling some quantities in histograms only takes a few lines of code,
in simple cases, and can be done with a single iteration over the input data.
For this task, which is very common and therefore used as an example here,
the main building blocks are the \Filter{} node, which only considers a subset
of the input data for which certain requirements are fulfilled,
the \HistoOneD{} node, which fills the values for a column into a histogram,
and the \Define{} node.
The latter allows to add a new column to the dataset, with its values calculated
from those of other columns, and can be constructed from a \CC{} code string
that may contain variables with the same names as existing columns,
or a \CC{} callable and a list of input columns this takes as arguments.

While in principle any derived column could be written as a single function
of the original inputs, the \Define{} nodes calculate their value once per
entry, and thus provide a form of caching that is essential for runtime
performance.
It is clear that in more complex analyses these \Define{} columns will contain
a significant part of the logic, and often relatively complex code.
When using \RDF{} directly from \CC{} this results in a splitting of the
analysis logic between helper function definitions and the construction of the
\RDF{} graph, or long and possibly repeated lambda functions; from python such
functions can either be implemented in \CC{} in a separate source file, or
JIT-compiled with numba from the same source file.
In all three cases it is challenging to keep the book-keeping of indices and
the corresponding arrays readable for code that is maximally performant,
especially with formats like NanoAOD, where collections are stored as a
structure of arrays, but the grouping of these arrays is partially based on
only naming conventions, and the stored columns may differ between files that
are analysed together.

Bamboo eliminates the need for writing such helper functions by providing a
higher-level interface that automates the construction of \Define{} nodes in
the \RDF{} graph, and that generates helper functions as needed: instead of
using the individual columns directly, it constructs an event view object,
with attributes that correspond to the collections and objects stored as
separate columns.
All of these are placeholder objects that behave like the value type of
the column or object they represent, \eg for collections a list interface with
random access and a length placeholder, and for objects an attribute for each
column.
This allows the user code to build up expressions for derived columns
piece by piece, by composing basic operations on the event view,
similarly as could be done in an object-oriented framework that processes
one record at a time.
The use of decorated placeholder objects to build up execution graphs is
analogous to machine learning frameworks like Theano~\cite{theano}
and Tensorflow~\cite{tensorflow2015-whitepaper}, but for dataset columns
instead of multi-dimensional arrays, and based on \RDF{} and cling.

The main difference with writing the event loop code directly is that
intra-event control flow in the execution graph is not introduced
by control flow in the python code written by the user,
but by calling helper functions that insert basic operations.
Many of these are commonly used higher-order functions from functional-style
programming, for instance \emph{select}, which takes a list and predicate,
and returns a new list with the elements for which the predicate is true;
see Figure~\ref{fig:fragmentOps} for an illustration of the typical code style.
Most of the information is then contained in the function arguments, which
represent predicates or transformations on one or more collection elements.
Inspiration for the higher-level functions was drawn from LoKi~\cite{LoKi}.

The example in Figure~\ref{fig:fragmentOps} also highlights another feature
of the event view and placeholder classes: the \texttt{p4} attribute is not
stored in the input dataset, but defined to automatically construct
a four-dimensional momentum vector from the stored components.
References between objects in different collections, which are stored as indices
in the NanoAOD file, are added to the proxy classes in a similar way.

\begin{figure}
  {\small\input{viewopsdeco.tex}}%
  \caption{Example bamboo code to select reconstructed muons that pass some
  selection criteria, construct a list of dimuon candidates with opposite charge,
  and calculate the mass of the first candidate.}%
  \label{fig:fragmentOps}%
\end{figure}

Together with the data-derived event view, the set of helper methods define
an embedded domain-specific language for constructing column expressions.
This approach has a number of advantages: the user-defined expressions are
restricted to the set of supported operations by the python interpreter,
the main analysis logic can easily be kept together in the same source file,
it is very compact, and it can use the full power of python, with
list comprehension, loops, and helper functions and classes, to construct
similar expressions without repeated code.

The implementation defines an interface implemented by about twenty concrete
classes to represent all such expressions, and about fifteen base placeholder
or proxy types that wrap an expression, out of which five are only needed
for automatic systematic variations, which are discussed in detail below.
The concrete event and collection view classes are dynamically generated
as subclasses with attributes based on the columns found in the tree,
using the \texttt{type} builtin.
Instances of the classes that represent operations are considered immutable
once they are fully constructed.
For each of them a hash is calculated that is unique for the expression they
represent, which allows for fast comparisons and the identification of
common subexpressions.
When a derived column is used in a \Filter{} or \HistoOneD{} node, all
subexpressions that should be precalculated for reuse, based on heuristics
of which operations are expensive and annotations provided by the user,
are automatically added as JIT-compiled \Define{} nodes in the \RDF{} graph.

\section{Selections, plots, and systematic uncertainties}
\label{sec:selectionplotsyst}

For the construction of \Filter{} and \HistoOneD{} nodes in the \RDF{} graph
a thin wrapper to insert \Define{} nodes would be sufficient, but given
the central role played by the different stages of selection on the input data,
bamboo opts for using its \Filter{} wrapper, the \Selection{} class, to include
some additional information, most importantly a per-event weight column.
Since many corrections are applied to simulated data as weights that depend
on per-event quantities, to reshape the corresponding distributions to match
those observed in data, it is natural to define a subset of the input data
and the corresponding weight by adding selection criteria (cuts) and weight
factors in steps, as more information is exploited to obtain a smaller subset.
The main method to construct such \Selection{} instances is by calling the
\emph{refine} method on an existing instance, with optional lists of cuts and
weights to add.

Filling a histogram with a distribution requires the columns with the quantity
of interest, a binning specification, the subsample to consider, and optionally
a per-event weight.
The latter two define a \Selection{}, so a \Plot{} instance --- the \HistoOneD{}
wrapper --- is constructed from a (derived) column expression, a binning
specification, and a \Selection{} instance.
Both \Selection{} and \Plot{} instances need to have a unique name, which is
necessary in the second case as a name for the histogram, and in both cases
useful to define columns with descriptive names, and for debugging printout.
The \Plot{} class can also hold a set of additional options to influence
the graphical display of the histogram, \eg colours, styles, and labels,
such that all the code to add the histograms for a stack plot, and its display
options, are kept together.
Figure~\ref{fig:fragmentPlotSelection} shows a typical example of user code
to define \Plot{} and \Selection{} instances.

\begin{figure}
  {\small\input{plotselection.tex}}%
  \caption{Example bamboo code to define \Plot{} and \Selection{} instances,
  see Section~\ref{sec:selectionplotsyst} for more details}%
  \label{fig:fragmentPlotSelection}%
\end{figure}

The hierarchy of \Selection{} objects that accumulates selection criteria and
weights not only maps very naturally on the \RDF{} graph, it also allows for
a simple generic implementation for what are often called data-driven background
estimates, where a disjoint data subsample is used to model a background
contribution, with a global or per-event transfer factor measured in yet another
subsample or simulation, as a weight, corresponding to the probability ratio
between the two selections.
This is implemented through a \Selection{} subclass that keeps not only the
main \Selection{}, but also one or more \Selection{} instances for such
backgrounds, with inverted selections and different weights: since the
\emph{refine} method of this subclass adapts all of the selection, no changes
elsewhere in the analysis code are required to include such a background
estimate.
In this way it is also easily possible to fill the histograms for different background
estimation strategies all at once to compare them.

\subsection*{Automatic systematic uncertainty variations}

In almost all realistic analysis use cases, the propagation of the uncertainty
on some measured quantities, or on the agreement between data and simulation,
is an important task, and a major source of complexity of the framework and
analysis code.
For the case of histograms that was considered above, the implication is that
for every histogram produced for a sample, \(1+N\) histograms now need to be
produced, each of the alternative versions with some changes in the weights,
selection criteria, binning variables, or several of these.
A major advantage of the domain-specific language used by bamboo is that this
can be done without changes to the analysis code: all that is required is
marking some inputs as having alternative values.
When creating \Selection{} and \Plot{} objects, the framework searches for such
marked inputs, and generates the \RDF{} graph to produce the necessary
additional histograms.
In case an input with variations is used in a selection criterion, a new branch
of the \RDF{} graph is created, where most of the nodes that are later added to
the main branch will need to be duplicated, so the size of the \RDF{} graph
may increase significantly.
If a variation only affects the weight and binning variable, only a few nodes
need to be added.
While this is foreseen to become directly supported in \RDF{}, the efficient
implementation of systematic variations, both code-wise and in runtime
performance, is currently one of the most compelling features of bamboo.

\section{\CC{} extensions}
\label{sec:cpptools}

In addition to the specific code described above, typical analyses also need
some building blocks that can more efficiently be implemented directly in
\CC{}.
Typical examples include loading additional weights or efficiency corrections
from a file, inference of multivariate classifier using a framework for
machine learning, the calculation of variations of some branch values,
additional reconstruction based on stored event quantities \etc{}
Since many of these can be widely shared across analyses, and most of the
current bamboo users are analysing CMS NanoAOD, \CC{} implementations
for several commonly used tasks are bundled.

Currently most efficiency corrections and weights are stored in a common
JSON format that was developed for an earlier framework, for which
a \CC{} reader class is included.
A module to reapply jet energy corrections, and to calculate variations of
the reconstructed jet energy that correspond to the uncertainty on
the jet energy correction and resolution, based on the values stored in the
NanoAOD, is also provided, including the propagation of these variations to
the reconstructed missing transverse momentum.
This is a nontrivial and crucial ingredient for producing results that
include all relevant systematic uncertainties based on the centrally produced
samples --- in many other frameworks for NanoAOD analysis an additional
processing step is needed, which stores the variations as columns in the file.
The possibility to apply, and easily change, corrections to the centrally
produced samples makes bamboo particularly suited to provide feedback about
new corrections and recipes, and to derive these, provided that the necessary
inputs are stored.

For the evaluation of multivariate classifiers using external machine learning
frameworks, a set of thin wrappers are provided with a very similar interface:
a class that is constructed with the path of the weights file and additional
information, if needed, with a method that takes the values for the features
for one example, and returns the value or values of the output classifier.
Such classes are currently available for TMVA~\cite{Hocker:2007ht} (through the
\texttt{TMVA::Experimental::RReader} class included in ROOT),
lwtnn~\cite{daniel_hay_guest_2020_4310003},
Tensorflow~\cite{tensorflow2015-whitepaper} (through the C API),
PyTorch~\cite{NEURIPS2019_9015} (using TorchScript), and
ONNX-Runtime~\cite{onnxruntime}.
Most of these support multiple input nodes, with potentially different types,
and are thread-safe, such that implicit multithreading~\cite{Piparo:2017hcq}
can be used.
The performance may in the future be improved by evaluating on a batch instead
of a single example at a time.

Technically, all the above follow a similar scheme: they are compiled
and linked as shared libraries, which are dynamically added to the ROOT runtime,
together with the corresponding headers.
As much configuration as possible is passed directly to the class constructor,
but in some cases the automatic python bindings provided by PyROOT are used
to pass additional settings by calling member methods on an instance from
the python code.
The same mechanism is used to implement unit tests that are integrated with
the pytest-based bamboo test suite, and PyROOT introspection is essential for
finding the return types where \CC{} methods are called.

The \CC{} template functions that implement basic functional algorithms are
defined in a header, and passed to ROOT for JIT compilation.
Many of these are range versions of STL algorithms, \eg the \emph{select}
algorithm used as an example before is essentially \texttt{std::copy\_if}
on a list of indices, so they may be changed to use, or be replaced by,
their standard counterparts in \CC{}20.

\section{Task management}
\label{sec:framework}

The bamboo framework is designed around the usually most resource-intensive task
of applying selections, calculating derived quantities, and filling histograms
and skimmed datasets.
Since these are performed by the \RDF{} event loop based on the graph built by
the user code, which is presented through the uniform interface of the
\texttt{definePlots} and \texttt{defineSkimSelection} methods that the user
module should implement, the framework can provide a shared implementation
for processing multiple samples, locally or on a batch system.

The list of samples are specified in a YAML configuration file that is passed
to the \texttt{bambooRun} command-line script together with the name of the
user analysis module.
Each sample entry needs to have a unique name, which is used as a name
for the output file, and specify how a list of files can be retrieved:
this can be a list of paths or URLs in the configuration file, the path of
a text file that contains these, or a query string for a sample management
database, \eg the CMS DAS system~\cite{Ball:2011zz}.
In the latter case it is also possible to keep a local disk cache of the query
results, to avoid executing them at every run.

The two actions performed by \texttt{bambooRun} are then obtaining a ROOT file
with histograms or skimmed TTrees for each sample, and postprocessing these to
obtain \eg combined plots.
The latter can also be run separately, such that changes that do not require
recreating the results files, \eg layout changes, or different labels, can be
applied in little time.
How the former is organised, is decided based on the additional arguments:
by default, the samples are processed sequentially.
The most common option is using a batch system to submit one or more jobs
that each process one sample, or part of a sample --- the splitting is then
defined in the YAML configuration file --- with the
\texttt{-{}-distributed=driver} option.
The commands for the individual tasks are almost identical, but with
\texttt{-{}-distributed=worker} and some additions to specify which sample and
files to process.
Since batch jobs may fail for various reasons, the driver mode checks
periodically for the status of all the jobs, merges the results for split
samples, and prints out information about failed jobs, including a command
to resubmit those after solving the problem, until all of the tracked jobs
have finished.
A \texttt{-{}-distributed=finalize} mode is also provided, which checks for
the per-job results, and performs any remaining merging if some jobs were
manually resubmitted.
Both the slurm and HTCondor batch submission systems are supported,
with the option to specify some site-specific configuration settings.
A plugin architecture is used, such that support for other batch systems can
be added with limited code changes.
A recent addition to \RDF{} is the possibility to trigger the execution of
multiple graphs in parallel from the same process.
This can be accommodated as an additional distributed mode without requiring
changes to the user code.

The postprocessing step may perform additional actions on the histograms and
skimmed datasets that are produced by the \RDF{} event loop, and stored in
ROOT files.
For skims nothing is done by default, while for histograms stack plots are made
using the standalone plotIt tool: since the sample and plot list are already
known, the only pieces of information that need to be added to the bamboo
configuration file for this are the cross-sections for the simulated samples,
the corresponding data luminosity, and the number of generated events in each
simulated sample.
The latter can also be calculated automatically based on the in-file metadata
for NanoAOD.

\section{Future directions}
\label{sec:futuredirections}

The bamboo framework already offers many options for customisation beyond
the most common workflows.
An example that is actively used is custom postprocessing: the default action is
to produce stack plots for one-dimensional histograms and ignores other outputs,
but it only takes a few lines of code to also make plots for two-dimensional
histograms, convert and rename all outputs, or register produced skims with
a database, since all the necessary information is readily available.
Some illustrations of processing other data formats than NanoAOD can be found
in a repository with examples on CMS and ATLAS open
data~\cite{bamboo_opendata_examples}, which are based on the corresponding
\RDF{} tutorials.
These also show the potential for analyses with less complexity than
measurements with the latest datasets and corrections:
for studies on open data, sensitivity studies on simulation, and
phenomenological analyses, the analysis code may in many cases fit
in a few screens.

An interesting side-effect of the embedded domain-specific language and its
implementation is that, although it maps very naturally to the \RDF{}
execution graphs, it avoids most direct calls of user code to ROOT and \RDF{}
directly.
This has also been observed while implementing a backend that does not build
the \RDF{} graph, but generates source code for a standalone \RDF{}-based \CC{}
program that does this: the dynamic features provided by ROOT are in that case
only needed for loading libraries and headers, and introspection through PyROOT,
and for directly passing configuration options to \CC{} classes, which is
only done in a few places.
That opens up possibilities beyond the current paradigm of batch-processing
large data samples and producing many thousands of histograms at once, but
instead caching results from previous runs, and only rerunning what is needed,
or passing queries to a dedicated system that optimizes the caching of columns
and datasets --- the unique hash of each derived column is currently an
implementation detail, but it could become an essential element in such a system.

\section{Conclusion}
\label{sec:conclusion}

The bamboo framework provides a convenient way to analyse
collision data, especially when using flat TTree formats like CMS NanoAOD.
It separates the user code, which is kept as simple, compact, and readable as possible
through an embedded domain-specific language, from its efficient execution,
which takes advantage of the possibilities provided by \RDF{} and cling.
The latter is currently optimised for processing on batch systems,
but the architecture allows for the different components to evolve separately
and thus adapt to evolutions of the computing environment.

\section*{Acknowledgements}

Bamboo would not have existed without the support and horizon provided by
the three-year Charg\'e de recherches mandate that the author
was awarded by the F.R.S.-FNRS in 2018.
It could also not have reached the current level of quality and maturity
without the support from the UCLouvain CP3 CMS team, in particular
S\'ebastien Wertz, Khawla Jaffel, Florian Bury, and
Gourab Saha (Saha Institute of Nuclear Physics),
who took the risk of adopting it early on, and have provided
invaluable feedback, ideas, suggestions, and help for the development.
Discussions with the ROOT team, in particular Enrico Guiraud,
and the participants of the PyHEP2019 workshop and several
meetings organised by the HEP software foundation, have also led to many
new insights and ideas for improvements.


\begin{thebibliography}{17}

\bibitem{Rizzi:2019rsi}
A.~Rizzi, G.~Petrucciani, M.~Peruzzi (CMS collaboration),
  \emph{{A further reduction in CMS event data for analysis: the NANOAOD format}},
  in Proceedings of the 23rd International Conference on Computing in High Energy and Nuclear Physics (CHEP 2018),
  \emph{EPJ Web of Conferences}, \textbf{214} (2019), 06021,
  \href{https://doi.org/10.1051/epjconf/201921406021}{10.1051/epjconf/201921406021}

\bibitem{enrico_guiraud_2017_260230}
E.~Guiraud, A.~Naumann, D.~Piparo,
  \emph{{TDataFrame: functional chains for ROOT data analyses}}, v1.0 (2017),
  \href{https://doi.org/10.5281/zenodo.260230}{10.5281/zenodo.260230}

\bibitem{Brun:1997pa}
  R.~Brun, F.~Rademakers,
  \emph{ROOT: An object oriented data analysis framework},
  Nucl. Instrum. Meth. A \textbf{389} (1997), 81--86

\bibitem{fons_rademakers_2020_3895852}
  F.~Rademakers et~al.,
  \emph{root-project/root}, v6.20/06 (2020),
  \href{https://doi.org/10.5281/zenodo.3895852}{10.5281/zenodo.3895852}

\bibitem{Sekmen:2020vph}
S.~Sekmen, P.~Gras, L.~Gray, B.~Krikler, J.~Pivarski, H.B. Prosper, A.~Rizzi, G.~Unel, G.~Watts,
  \emph{Analysis Description Languages for the LHC},
  PoS \textbf{LHCP2020}, 065 (2021),
  \href{https://doi.org/10.22323/1.382.0065}{10.22323/1.382.0065},
  \href{https://arxiv.org/abs/2011.01950}{arXiv:2011.01950}

\bibitem{cling}
V.~Vassilev, P.~Canal, A.~Naumann, L.~Moneta, P.~Russo,
  \emph{{Cling} -- The New Interactive Interpreter for {ROOT} 6},
  Journal of Physics: Conference Series \textbf{396} (IOP, 2021) 5, 052071,
  \href{https://doi.org/10.1088/1742-6596/396/5/052071}{10.1088/1742-6596/396/5/052071},

\bibitem{lavrijsen2016high}
W.T. Lavrijsen, A.~Dutta, \emph{High-performance Python-C++ bindings with PyPy
  and Cling}, in \emph{6th Workshop on Python for High-Performance and
  Scientific Computing (PyHPC)} (IEEE, 2016), 27--35,
  \href{https://cppyy.readthedocs.io/en/latest/}{https://cppyy.readthedocs.io/en/latest/}

\bibitem{theano}
J.~Bergstra, O.~Breuleux, F.~Bastien, P.~Lamblin, R.~Pascanu, G.~Desjardins, J.~Turian, D.~Warde-Farley, Y.~Bengio,
  \emph{Theano: A CPU and GPU math compiler in Python},
  in Proceedings of the 9th Python in Science conference (2010),
  \href{https://doi.org/10.25080/Majora-92bf1922-003}{10.25080/Majora-92bf1922-003}

\bibitem{tensorflow2015-whitepaper}
M.~Abadi et~al.,
  \emph{{TensorFlow}: Large-scale machine learning on heterogeneous systems}
  (2015), software available from \href{https://www.tensorflow.org/}{tensorflow.org}

\bibitem{LoKi}
I.~Belyaev, \emph{{LOKI: Smart \& Friendly C++ Physics Analysis Toolkit}},
  \href{http://cern.ch/lhcb-comp/Analysis/Loki}{http://cern.ch/lhcb-comp/Analysis/Loki}

\bibitem{Hocker:2007ht}
A.~Hoecker, P.~Speckmayer, J.~Stelzer, J.~Therhaag, E.~von Toerne, H.~Voss,
  \emph{TMVA: Toolkit for Multivariate Data Analysis},
  PoS \textbf{ACAT}, 040 (2007), CERN-OPEN-2007-007,
  \href{https://arxiv.org/abs/physics/0703039}{arXiv:physics/0703039}

\bibitem{daniel_hay_guest_2020_4310003}
D.H. Guest, J.W. Smith, M.~Paganini, M.~Kagan, M.~Lanfermann, A.~Krasznahorkay, D.E. Marley, A.~Ghosh, B.~Huth,
  \emph{lwtnn/lwtnn}: version 2.11.1 (2020),
  \href{https://doi.org/10.5281/zenodo.4310003}{10.5281/zenodo.4310003}

\bibitem{NEURIPS2019_9015}
A.~Paszke et~al.,
  \emph{PyTorch: An Imperative Style, High-Performance Deep Learning Library},
  in \emph{Advances in Neural Information Processing Systems 32}
  (2019), pp. 8024--8035,
  \href{https://arxiv.org/abs/1912.01703}{arXiv:1912.01703}

\bibitem{onnxruntime}
\emph{{ONNX Runtime: Optimize and Accelerate Machine Learning Inferencing and
  Training}}, \url{https://www.onnxruntime.ai}

\bibitem{Piparo:2017hcq}
D.~Piparo, E.~Tejedor, E.~Guiraud, G.~Ganis, P.~Mato, L.~Moneta, X.~Valls~Pla, P.~Canal,
  \emph{Expressing Parallelism with ROOT},
  J. Phys. Conf. Ser. \textbf{898} (2017) 7, 072022,
  FERMILAB-CONF-16-738-CD,
  \href{https://doi.org/10.1088/1742-6596/898/7/072022}{10.1088/1742-6596/898/7/072022}

\bibitem{Ball:2011zz}
G.~Ball, V.~Kuznetsov, D.~Evans, S.~Metson,
  \emph{Data Aggregation System: A system for information retrieval on demand over relational and non-relational distributed data sources},
  J. Phys. Conf. Ser. \textbf{331} (2011), 042029,
  FERMILAB-CONF-11-874-CMS,
  \href{https://doi.org/10.1088/1742-6596/331/4/042029}{10.1088/1742-6596/331/4/042029}

\bibitem{bamboo_opendata_examples}
P.~David, \emph{bamboo {Open Data} examples},
  \href{https://github.com/pieterdavid/bamboo-opendata-examples}{pieterdavid/bamboo-opendata-examples}

\end{thebibliography}

\end{document}